\documentclass[aps,twocolumn,pra,superscript,floatfix,superscriptaddress,showpacs]{revtex4-1}

\usepackage{epsf}
\usepackage{epsfig}
\usepackage{graphicx}
\usepackage{dcolumn}
\usepackage{braket}
\usepackage{bm}
\usepackage{amsfonts}
\usepackage{amsmath}
\usepackage{amssymb}
\usepackage{color,soul}
\usepackage{wasysym}
\usepackage{mathrsfs}
\usepackage{natbib}
\usepackage{float}
\usepackage{multirow}
\usepackage[colorlinks=true,%
            linkcolor=blue,%
            urlcolor=blue,%
            citecolor=blue,%
            filecolor=blue,%
            bookmarksopen=true,%
            pdfauthor={J. P. Ramos-Andrade},%
            pdftitle={BIC},%
            pdfsubject={BIC_Manuscript},%
            pdfpagemode=UseOutlines]{hyperref}

\newcommand{\la}{\langle}
\newcommand{\ra}{\rangle}
\newcommand{\ve}{\varepsilon}
\newcommand{\up}{\uparrow}
\newcommand{\down}{\downarrow}
\newcommand{\drm}{\mathrm{d}}

\definecolor{otrogreen}{rgb}{0, 0.7, 0.}
\definecolor{antiquefuchsia}{rgb}{0.57, 0.36, 0.51}
\definecolor{amethyst}{rgb}{0.6, 0.4, 0.8}

\begin{document}

\title{Fano-Rashba effect in the presence of Majorana bound states}
\author{B. Grez}
\affiliation{Departament of Physics, Federico Santa Mar\'ia Technical University, Valpara\'iso, Chile}
\author{J. P. Ramos-Andrade}
\affiliation{Department of Physics, University of Antofagasta, Av. Angamos 601, Antofagasta, Chile.}
\author{P.\ A.\ Orellana}
\affiliation{Departament of Physics, Federico Santa Mar\'ia Technical University, Valpara\'iso, Chile}
\

\begin{abstract}
In this paper, we investigate the influence of Majorana bound states on the Fano-Rashba effect in a two-channel Fano-Anderson model. Employing Green's function formalism and the equation of motion method, we compute the transmission through the quantum dot and the density of states. Our analysis reveals that the Majorana bound states, localized at the ends of the topological superconductor nanowire, penetrate into the quantum dot, thereby altering the interference pattern in the electronic transmission profile through it, resulting from their interaction with the bound states in the continuum. Furthermore, we explore the robustness of the bound state in the continuum concerning its connection to Majorana bound states and the energy induced by the magnetic field. We posit that our findings contribute to a deeper comprehension of the Fano-Rashba effect in a two-channel quantum dot coupled to a topological superconducting nanowire. 
\end{abstract}

\maketitle

\section{Introduction}

The Fano-Rashba effect in quantum dots (QDs) has been previously explored in literature \cite{hanson2007spins,holleitner2001coherent,holleitner2002probing,orellana2003transport,ramos2014bound,van2002electron}. This phenomenon occurs due to the spin-orbit Rashba coupling, which allows for mixing the spin degree of freedom of propagating electrons and creates an additional source of interference \cite{hsu2016bound}. The latter can be understood as follows: when an electron with a specific spin $\sigma$ enters the QD, it may maintain or invert its spin, resulting in destructive interference between the two possible paths, leading to the emergence of the named Fano effect \cite{fano1961effects}. Additionally, we previously reported forming a bound state in the continuum (BIC) within a generic two-channel Fano-Anderson model, i.e., electronic or bosonic \cite{Grez2022bound}. The mechanism occurs when the mixing coupling between the channels (spin channels, for instance) equals the direct intra-channel coupling, obtaining a BIC formed in the QD. On the other hand, there has been much interest in investigating BICs in the last few years since several implementations can be performed based on it. The BICs have energies within the continuum band, such as the conduction or radiation band. However, they do not overlap with it and do not decay as a primary consequence. These were predicted for the first time during the early days of quantum mechanics by von Neumann and Wigner \cite{vonNeumann-Wigner}. The first observation of BICs was achieved in photonics systems and then in photonics waveguides \cite{hsu2013observation}. More recently, they have been detected in sound waves and even used in technology, such as designing lasers based on BICs \cite{Marinica2008bound,Plotnik2011experimental}. 

In condensed matter physics, a significant focus has been on the study of topological superconductor nanowires (TSCs) in the last decade \cite{Sato_2017}. The presence of unique fermionic quasiparticles, known as Majorana bound states (MBSs), has been predicted to exist in these systems. These quasiparticles are believed to be their own anti-quasiparticles, similar to Majorana fermions \cite{majorana1937teoria}, and are commonly predicted to be found localized at the edges of TSCs. One of the most studied systems corresponds to a one-dimensional (1D) TSC modeled by Kitaev \cite{kitaev2001unpaired}, where both ends are predicted to host MBSs. The interest in studying such systems is because the MBSs satisfy non-Abelian statistics, and therefore, they are seen as potential candidates to implement fault-tolerant quantum computing technology \cite{alicea2016exponential}.

The first physical realization of a 1D topological superconductor was achieved in 2012 by Mourik and collaborators, who announced the emergence of zero-bias anomalies in the conductance as a signature of the presence of Majorana bound states. Following this milestone, many experiments based on zero-bias anomalies in transport properties measurements through source-drain leads have been performed \cite{deng2012anomalous,das2012zero,lee2012zero,finck2013anomalous,churchill2013superconductor}. Nowadays, it is known that these anomalies are not always reliable evidence of MBSs, leading to the necessity of implementing additional protocols, for instance, for identifying the non-locality of MBSs exploring current shot noise correlations \cite{Manousakis2020weak}, and devising custom-made experimental protocols that allow e.g., performing simultaneous tunneling and Coulomb blockade spectroscopy measurements within the same device, to rule out MBSs detection ambiguities \cite{valentini2022majorana}. Recently, in Ref. \cite{Aghaee2023InAs}, it was reported hybrid devices pass a gap protocol designed to identify the topological superconducting phase of the device.
\begin{figure*}[t]
\includegraphics[width=\textwidth]{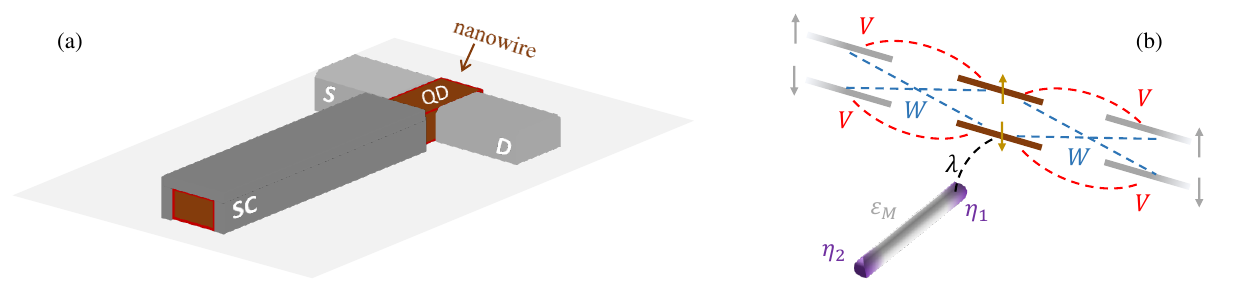}
\centering
\caption{(a) Experimental view of the setup: a QD coupled to source ($S$) and drain ($D$) leads and connected to a TSC. (b) Schematic view of the setup: the two channels of the QD coupled to source/drain leads channels. The parameter $V$ (red dashed lines) denotes the intra-channel coupling, while the parameter $W$ (blue dashed lines) corresponds to the inter-channel coupling. The MBSs $\eta_{1}$ and $\eta_{2}$ are placed at the ends of the TSC (purple zones). The parameter $\lambda$ denotes the coupling between one of the QD's channels and $\eta_{1}$, while $\eta_{1}$ and $\eta_{2}$ interact between them with tunneling coupling $\ve_{M}$.}\label{fig1}
\end{figure*}

The interplay between trivial BICs and MBSs has been previously addressed in systems considering TSCs coupled to a single QD \cite{ricco2016decay,RamosAndrade2019FanoMajorana,RamosAndrade2020majorana,Zhang2022QD} and to multiple QDs \cite{guessi2017encrypting,garrido2023bound,Zambrano2018bound}. Nevertheless, the latter interplay has not been so far in a QD-TSC system with a Rashba-like coupling between leads and QD. In this article, we investigate the  Fano-Rashba effect in a single QD coupled to a TSC nanowire hosting two MBSs at its ends. We use the two-channel Fano-Anderson model to describe the system, including Rashba interaction and coupling to the TSC. To solve this problem, we employ Green's function formalism and the equation of motion procedure. Our calculations of the transmission spectrum and density of states (DOS) in the QD reveal the emergence of Fano antiresonances caused by the Rashba interaction and the coupling with the TSC and also reveal the conditions of the parameters to obtain a BIC in our system robustly.

The paper is organized as follows. In Sec.\ \ref{model}, we present the model and formalism considered to address the problem, while in Sec.\ \ref{Results}, the results are displayed, and the related discussion is performed. The final remarks are in Sec.\ \ref{summary}.

\section{Model and Formalism}\label{model}

The system under study corresponds to a two-channel Anderson model, considering those as the two spin channels in a QD, where one of these channels is coupled with a 1D TSC hosting MBSs at its ends. We show the system schematically in Fig.\ \ref{fig1}, which is described through an effective low-energy Hamiltonian $H$ in the form
\begin{equation}\label{Hinicial}
H=H_{\text{dot}}+H_{\text{L}}+H_{\text{dot-L}}+H_{\text{R}}+H_{\text{dot-M}}+H_{\text{M}}\,,
\end{equation}
where the first two terms on the rhs denote the two channels of the QD and the leads, respectively, given by
\begin{eqnarray}
    H_{\text{dot}}&=&\sum_{\sigma} \ve_{\sigma}d_{\sigma}^{\dag}d_{\sigma}\,,\\
    H_{\text{L}}&=&\sum_{\mathbf{k}, \sigma, \alpha}\ve_{\mathbf{k}, \sigma, \alpha}c_{\mathbf{k}, \sigma, \alpha}^\dag c_{\mathbf{k}, \sigma, \alpha}\,,
\end{eqnarray}
where $c_{\mathbf{k}, \sigma, \alpha}^{\dag}(c_{\mathbf{k}, \sigma, \alpha})$ creates (annihilates) an electron with momentum $\mathbf{k}$, spin $\sigma=\,\up$ or $\down$, and energy $\ve_{\mathbf{k}, \sigma, \alpha}$ in the lead $\alpha=S,D$, where $S$ and $D$ stands for source and drain leads, respectively. Also, $d_{\sigma}^{\dag}(d_{\sigma})$ creates (annihilates) an electron with energy $\ve_{\sigma}$ in the QD's level corresponding to the spin $\sigma$. The Hamiltonian terms $H_{\text{dot-L}}$ and $H_{\text{R}}$ correspond to the connection between the leads and the QD's levels, which are given by
\begin{eqnarray}
    H_{\text{dot-L}}&=&\sum_{\mathbf{k}, \sigma, \alpha}\left(V^*c_{\mathbf{k}, \sigma, \alpha}^\dag d_{\sigma} +Vd_{\sigma}^\dag c_{\mathbf{k}, \sigma, \alpha}\right)\,,\\
    H_{\text{R}}&=&\sum_{\mathbf{k}, \sigma, \sigma', \alpha}\left[Wd_{\sigma'}^\dag c_{\mathbf{k}, \sigma, \alpha}+W^*c_{\mathbf{k}, \sigma, \alpha}^\dag d_{\sigma'}\right]\sigma^x_{\sigma\sigma'},
\end{eqnarray}
where $V$ and $W$ are the couplings between the $\sigma$ channel of the lead $\alpha$ with the QD's channel $\sigma$, and $\sigma'\neq\sigma$, respectively. Note that $\sigma^{x}$ denotes the Pauli's matrix $x$. 

The last two terms in Eq.\ (\ref{Hinicial}) correspond to the MBSs and their connections with the lower channel (with spin $\sigma=\,\down$ in Fig.\ \ref{fig1}) of the QD. To write $H_{\text{dot-M}}$ and $H_{\text{M}}$ we consider each MBS operator $\eta_{1}$ and $\eta_{2}$, that satisfy both $\eta_{l}=\eta_{l}^{\dag}$ and $\{\eta_{l},\eta_{l'}\}=\delta_{l,l'}$ ($l=1,2$), as a superposition of regular fermionic operators $f=(\eta_{1}+i\eta_{2})/\sqrt{2}$ and $f^{\dag}=(\eta_{1}-i\eta_{2})/\sqrt{2}$, which satisfy the regular fermionic anticommutation relation. Accordingly, we express   
\begin{eqnarray}
    H_{\text{dot-M}}&=&\frac{\lambda}{\sqrt{2}}(d_{\down}-d_{\down}^{\dag})(f+f^{\dag})\,,\\
    H_{\text{M}}&=&\ve_{M}\left(f^{\dag}f-\frac{1}{2}\right)\,,
\end{eqnarray}
where $\ve_{M}$ is the tunneling coupling between the MBSs placed at opposite ends. Since we adopt Kitaev's model, $\ve_{M}$ decay exponentially with the length $L$ of the TSC, i.e. $\ve_{M}\propto\exp{(-L/\xi)}$, where $\xi$ is the superconducting coherence length.

We employ Green's function (GF) formalism to address the problem and to obtain the physical quantities of interest. In the time domain, the regular elements of the QD's retarded GF are defined by 
\begin{align}
    G_{\sigma,\sigma'}^{\text{r}}(t)&=\la\la d_{\sigma}(t);d_{\sigma'}^\dag(0)\ra\ra_{t}^{\text{r}}=-\frac{i}{\hbar}\Theta(t)\la\{d_{\sigma}(t),d_{\sigma'}^\dag(0)\}\ra\,, \\
    \mathcal{G}_{\sigma,\sigma'}^{\text{r}}(t)&=\la\la d_{\sigma}^\dag(t);d_{\sigma'}(0)\ra\ra_{t}^{\text{r}}=-\frac{i}{\hbar}\Theta(t)\la\{d_{\sigma}^{\dag}(t),d_{\sigma'}(0)\}\ra\,,
\end{align}
while the anomalous elements are defined by 
\begin{align}
    F_{\sigma,\sigma'}^{\text{r}}(t)&=\la\la d_{\sigma}^\dag(t);d_{\sigma'}^\dag(0)\ra\ra_{t}^{\text{r}}=-\frac{i}{\hbar}\Theta(t)\la\{d_{\sigma}^{\dag}(t),d_{\sigma'}^\dag(0)\}\ra\,, \\
    \mathcal{F}_{\sigma,\sigma'}^{\text{r}}(t)&=\la\la d_{\sigma}(t);d_{\sigma'}(0)\ra\ra_{t}^{\text{r}}=-\frac{i}{\hbar}\Theta(t)\la\{d_{\sigma}(t),d_{\sigma'}(0)\}\ra\,.
\end{align}

In terms of the GF, the transmission coefficient can be expressed by the Fisher-Lee relation as follows.

\begin{equation}\label{FisherLee}
    T(\ve)=\text{Tr}\{ \frak{G}^{\text{a}}(\ve)\Gamma^R \frak{G}^{\text{r}}(\ve)\Gamma^L\}\,,
\end{equation}
where $\frak{G}^{\text{r(a)}}(\ve)$ is the retarded(advanced) GF in energy domain, that satisfy $\left[\frak{G}^{\text{r}}(\ve)\right]^{\dag}=\frak{G}^{\text{a}}(\ve)$, and has the form 
\begin{equation}\label{greenmatriz}
     \frak{G}^{\text{r}}(\ve) = \left(\begin{array}{cccc}
        G_{\up,\up}^{\text{r}}(\ve) & G_{\up,\down}^{\text{r}}(\ve) & \mathcal{F}_{\up,\up}^{\text{r}}(\ve) & \mathcal{F}_{\up,\down}^{\text{r}}(\ve) \\
        G_{\down,\up}^{\text{r}}(\ve) & G_{\down,\down}^{\text{r}}(\ve) & \mathcal{F}_{\down,\up}^{\text{r}}(\ve) & \mathcal{F}_{\down,\down}^{\text{r}}(\ve) \\
        F_{\up,\up}^{\text{r}}(\ve)  & F_{\up,\down}^{\text{r}}(\ve)  & \mathcal{G}_{\up,\up}^{\text{r}}(\ve) & \mathcal{G}_{\up,\down}^{\text{r}}(\ve) \\
        F_{\down,\up}^{\text{r}}(\ve)  & F_{\down,\down}^{\text{r}}(\ve) & \mathcal{G}_{\down,\up}^{\text{r}}(\ve) & \mathcal{G}_{\down,\down}^{\text{r}}(\ve)
    \end{array} \right)\,.
\end{equation}
Besides, $\Gamma^{\alpha}$ is the energy-independent coupling matrix that connects the lead $\alpha$ with the QD, which is given by
\begin{equation}
    \Gamma^{\alpha}= \left(\begin{array}{cccc}
         \gamma(1+r^{2}) & 2\gamma r & 0 & 0 \\
         2\gamma r & \gamma(1+r^{2}) & 0 & 0 \\
         0 & 0 & \gamma(1+r^{2}) & 2\gamma r \\
         0 & 0 & 2\gamma r & \gamma(1+r^{2})
    \end{array} \right)\,,
\end{equation}
where we have defined the dimensionless parameter $r$ as $r=W/V$, and $\gamma=2\pi\rho_{0}\lvert V\lvert^2$, with $\rho_{0}$ as the constant leads' DOS within the wide-band approximation. 

In a previous work \cite{Grez2022bound}, it was shown that a generic (fermionic or bosonic) two-channel Fano-Anderson model supports true BICs for the symmetric coupling case, i.e., $W=V$ or $r=1$ in this work. This BIC is related to the antisymmetric state of the system. In order to study the robustness of the BIC in the presence of the MBSs, we perform a symmetrization through a unitary operator $U$ in the form
\begin{equation}
    U = \frac{1}{\sqrt{2}} \left(\begin{array}{cccc}
         1 & 1 & 0 & 0 \\
         1 & -1 & 0 & 0 \\
         0 & 0 & 1 & 1 \\
         0 & 0 & 1 & -1
    \end{array} \right)\,,
\end{equation}
separating the contribution of the symmetric/antisymmetric state to the physical quantities. Note that $U$ can be view as the following Kronecker product $U=I_{2}\otimes\mathcal{H}$, where $I_{2}$ and $\mathcal{H}$ are the $2\times 2$ identity matrix and the Hadamard matrix, respectively. Accordingly, Eq.\ (\ref{FisherLee}) transforms to

\begin{align}
    T(\ve) &= \text{Tr}\{U^{\dag} \frak{G}^{\text{a}}(\ve)UU^{\dag}\Gamma^R UU^{\dag} \frak{G}^{\text{r}}(\ve) UU^{\dag} \Gamma^L U\}\nonumber \\
    &= \text{Tr}\{ \tilde{\frak{G}}^{\text{a}}(\ve) \tilde{\Gamma}^{R} \tilde{\frak{G}}^{\text{r}}(\ve) \tilde{\Gamma}^{L}  \}\,,
\end{align}
where the transmission coefficient is expressed in terms of modified GFs $\tilde{G}^{\text{a(r)}}$, whose new elements are given by 

\begin{align}
    \tilde{\frak{B}}_{+,+}^{r}(\ve) &= \frac{1}{2} \left[\frak{B}_{\up,\up}^{r}(\ve) + \frak{B}_{\up,\down}^{r}(\ve) + \frak{B}_{\down,\up}^{r}(\ve) + \frak{B}_{\down,\down}^{r}(\ve)\right]\,
    \label{eq:greenpp}, \\
    \tilde{\frak{B}}_{-,-}^{r}(\ve) &= \frac{1}{2} \left[\frak{B}_{\up,\up}^{r}(\ve) - \frak{B}_{\up,\down}^{r}(\ve) - \frak{B}_{\down,\up}^{r}(\ve) + \frak{B}_{\down,\down}^{r}(\ve)\right]\,, \\
    \tilde{\frak{B}}_{+,-}^{r}(\ve) &= \frac{1}{2} \left[\frak{B}_{\up,\up}^{r}(\ve) - \frak{B}_{\up,\down}^{r}(\ve) + \frak{B}_{\down,\up}^{r}(\ve) - \frak{B}_{\down,\down}^{r}(\ve)\right]\,, \\
    \tilde{\frak{B}}_{-,+}^{r}(\ve) &= \frac{1}{2} \left[\frak{B}_{\up,\up}^{r}(\ve) + \frak{B}_{\up,\down}^{r}(\ve) - \frak{B}_{\down\up}^{r}(\ve) - \frak{B}_{\down,\down}^{r}(\ve)\right]\,
    \label{eq:greenmp},
\end{align}
where $\frak{B}$ stands for either $G,\mathcal{G},F$ or $\mathcal{F}$, and the modified coupling matrix $\tilde{\Gamma}$, given by
\begin{equation}
    \tilde{\Gamma}^{\alpha} = \left( \begin{array}{cccc}
         \gamma_{+}& 0 & 0 & 0 \\
         0 & \gamma_{-} & 0 & 0 \\
         0 & 0 & \gamma_{+} & 0 \\
         0 & 0 & 0 & \gamma_{-}
    \end{array} \right)\,,
\end{equation}
where we have defined $\gamma_{+}=\gamma(1+r)^{2}$ and $\gamma_{-}=\gamma(1-r)^{2}$ \cite{Grez2022bound}. Then, within the transformation, the transmission coefficient is given by

\begin{align}
    T(\ve)= \sum_{\nu,\nu'}& \left(  \lvert \tilde{G}_{\nu,\nu'}(\ve) \lvert^{2} \gamma_{\nu} \gamma_{\nu'} + \lvert \tilde{F}_{\nu,\nu'}(\ve) \lvert^{2} \gamma_{\nu} \gamma_{\nu'} + \right. \\
    & \left. \lvert \tilde{\mathcal{G}}_{\nu,\nu'}(\ve) \lvert^{2} \gamma_{\nu} \gamma_{\nu'} + \lvert \tilde{\mathcal{F}}_{\nu,\nu'}(\ve) \lvert^{2} \gamma_{\nu} \gamma_{\nu'} \right)\,,
\end{align}
where $\nu = +,-$.

\begin{figure*}[t!]
\centering
\includegraphics[width=.9\linewidth]{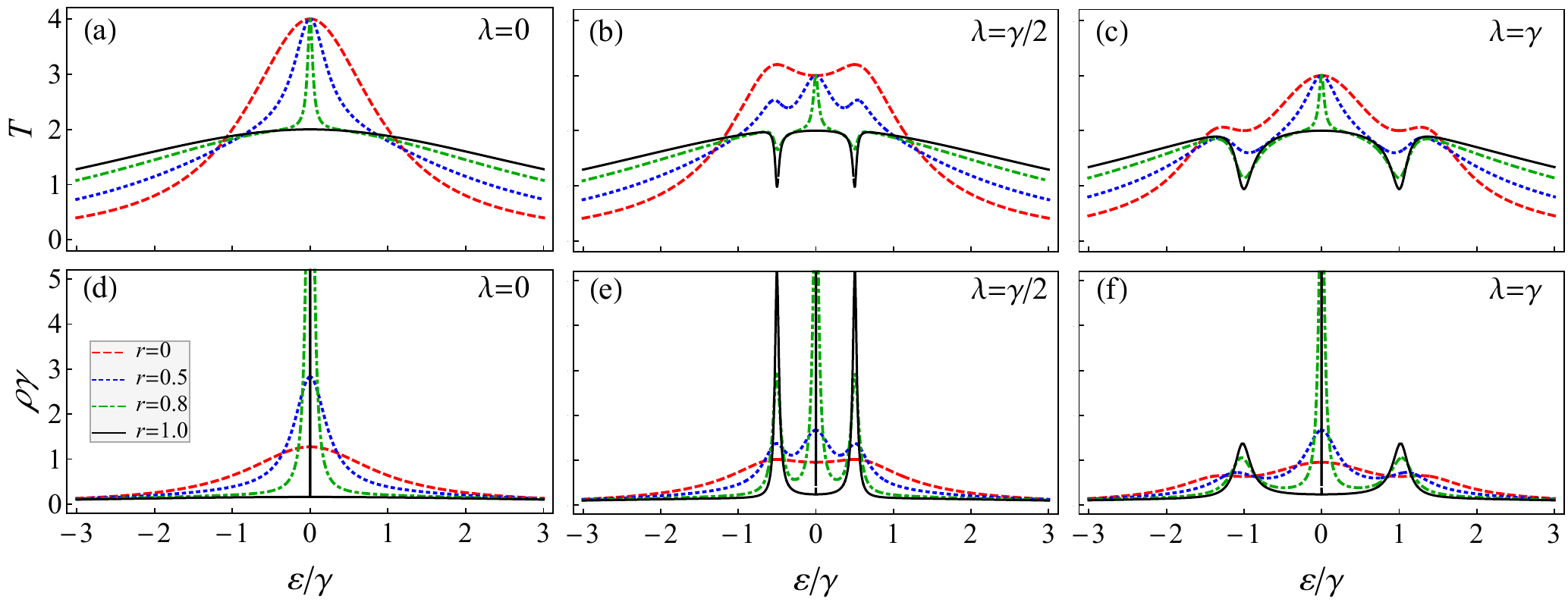}
\caption{Transmission probability $T$ (upper panels) and dimensionless density of states $\rho$ (lower panels), both as functions of the energy $\ve$ for fixed: [(a) and (d)] $\lambda=0$; [(b) and (e)] $\lambda=\gamma/2$; and [(c) and (f)] $\lambda=\gamma$. Different values of $r$ are considered, from $r=0$ to $r=1$.}\label{fig2}
\end{figure*}

\begin{figure}[h!]
\centering
\includegraphics[width=\linewidth]{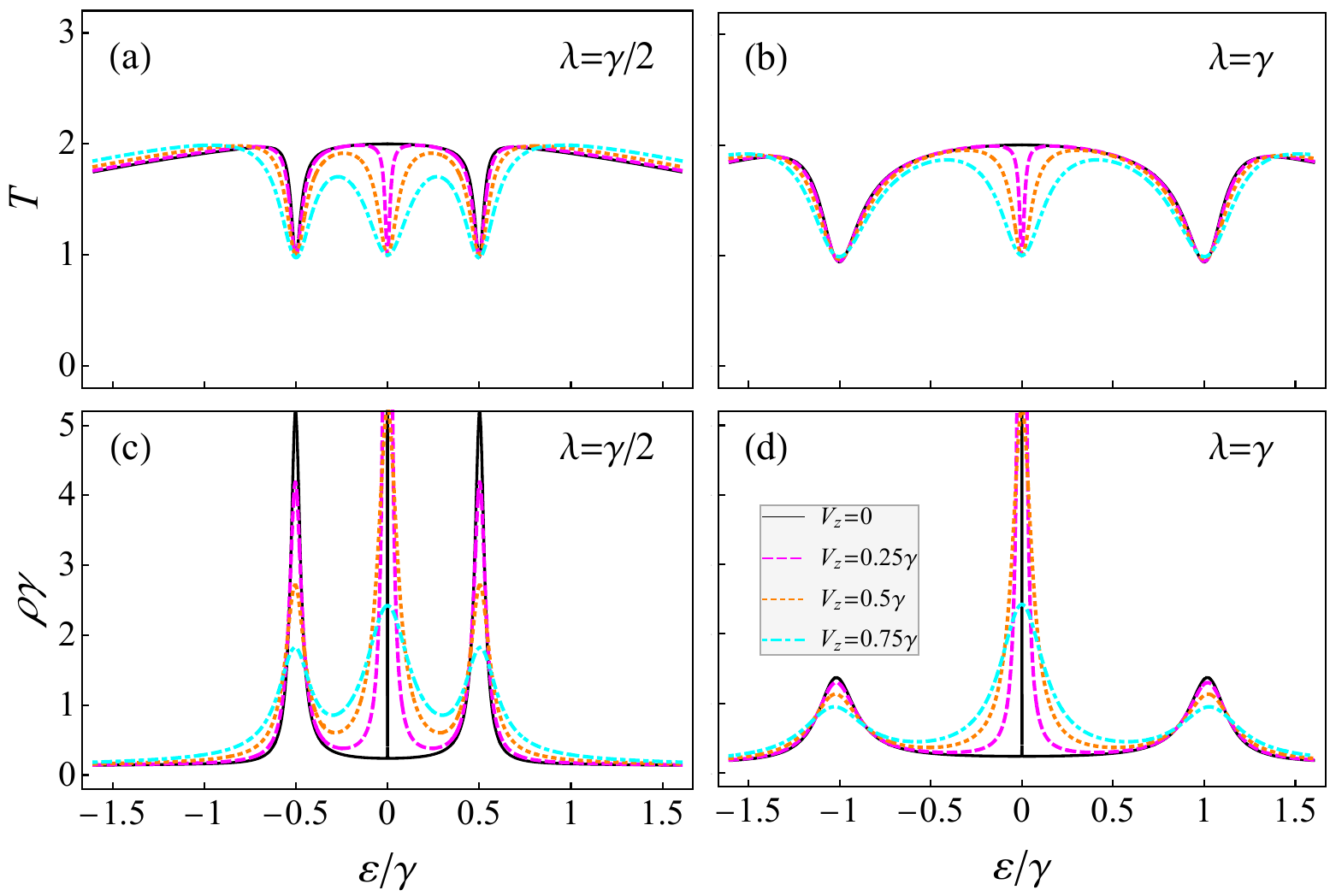}
\caption{Transmission probability $T$ (upper panels) and dimensionless density of states $\rho$ (lower panels) as functions of the energy $\ve$ using $r=1$ for fixed: [(a) and (c)] $\lambda=\gamma/2$; and [(b) and (d)] $\lambda=\gamma$. Different values of $V_{z}$ are considered, from $V_{z}=0$ to $V_{z}=0.75\gamma$.}\label{fig3}
\end{figure}

\section{Results}\label{Results}

In what follows, we present the obtained results using the parameter $\gamma$ as the energy unit of the system. In order to consider the relation between the parameters used with experiments, the values for $\gamma$ can be considered from a few to hundreds of meV. Since we consider the two channels as the spin-resolved channels in the QD, we include a magnetic field of magnitude $B$. Then, spin-resolved energy levels are given by $\ve_{\sigma}=\ve_{d}+\sigma_{\sigma\sigma}^{z}V_{z}$, where $V_{z}=g\mu_{\text{B}}B$ is the Zeeman's energy, being $\mu_{\text{B}}$ the Bohr magneton, and $g$ the Land\'e factor.  Note that $\sigma^{z}$ denotes Pauli’s matrix $z$.

Figure\ \ref{fig2} displays the transmission probability and the DOS as a function of the energy, considering different values of the Rashba-like coupling parameter expressed through the ratio $r=W/V$, from $r=0$ (red dashed line) to $r=1$ (black solid line), going through intermediate values (blue dotted and green dot-dashed lines). From Fig.\ \ref{fig2}(a) and Fig.\ \ref{fig2}(d), where the TSC is uncoupled of the QD's spin down channel, we can observe the evolution as $r$ increases of a wide resonance to a superposition of a broad and sharp resonance, obtaining a true BIC in the DOS for the case of $r=1$, consequently without projection in the transmission probability, exhibiting a half-maximum value at zero energy, as was reported in Ref.\ \cite{Grez2022bound}. Both transmission probability and DOS can be separated into two overlapped contributions with widths $\gamma_{+}$ (broad) and $\gamma_{-}$ (sharp).
It is worth mentioning that in this case, $T_{\text{max}}=4$ is due to the electron-hole redundancy in the channels included by the TSC. The effect of the connection of the MBSs is presented in panels Fig.\ \ref{fig2}(b), Fig.\ \ref{fig2}(c), Fig.\ \ref{fig2}(e), and Fig.\ \ref{fig2}(f) by using $\lambda\neq 0$. In this scenario, in the case of $r=1$, the true BIC remains robustly pinned at zero-energy regardless of the $\lambda$ value considered. On the other hand, two additional states are observed in panels FIG.\ \ref{fig2}(e) and FIG.\ \ref{fig2}(f), symmetrically placed at energies $\ve=\pm\lambda$, whose projections in the transmission probability are observed as non-vanishing antiresonances in panels Fig.\ \ref{fig2}(b) and Fig.\ \ref{fig2}(c). The width of the latter antiresonances is proportional to $\lambda^{2}/\gamma$. On the other hand, whenever $r\neq 0$, the transmission probability at zero-energy decreases up to $0.75T_{\text{max}}$. This can be understood as the maximum transmission halving of the antisymmetric transmission channel.

Here, we focus on the relevant case of $r=1$. Fi\-gu\-re\ \ref{fig3} shows the transmission probability and DOS as a function of the energy in the presence of the applied magnetic field, expressed utilizing different Zeeman energy $V_{z}$ values. The BIC observed at zero-energy for vanishing $V_{z}$ evolves to a finite-width resonance, i.e., a quasi-BIC, whenever $V_{z}\neq 0$. As $V_{z}$ increases, the quasi-BIC resonance becomes wider, and its amplitude decreases. The latter directly shortens the lifetime of propagating electrons into this state. When a magnetic field is applied ($V_{z}\neq 0$), the transmission probability spectrum exhibits an antiresonance around zero-energy, reaching a half-integer value regardless of both $V_{z}$ and $\lambda$ values. In this scenario, the transmission probability spectrum can be represented analytically as a convolution of a Breit-Wigner function and two Fano lines shapes, where the latter has an imaginary and dimensionless $ q$ parameter, known as the Fano-factor. The expression is given as 
\begin{equation}
 T(\varepsilon)\simeq\frac{4\gamma^2}{\varepsilon^2+4\gamma^2}  \frac{|\epsilon+q|^2}{\epsilon^2+1}\frac{|\xi+q|^2}{\xi^2+1},
\end{equation}
where $\epsilon=(|\varepsilon|-\lambda)/\eta_{\lambda}$ , $\xi=\varepsilon/\eta_z$ and $q=i/\sqrt{2}$. The parameters $\eta_{z}\propto V_{z}^{2}/\gamma$ and $\eta_{\lambda}\propto\lambda^{2}/\gamma$ are related to the widths of the central and lateral dips.
It is important to note that in the equation above, the last factor on the right-hand side corresponds to a Fano line shape that tends to unity without a magnetic field and when $\eta_{z}\rightarrow 0$. Similarly, an analytical analysis can be performed for the DOS, which can be expressed as a superposition of three Lorentzian line shapes.
\begin{equation}
 \rho(\varepsilon)\simeq\frac{1}{\pi}\left(\frac{2\gamma}{\varepsilon^2+4\gamma^2} +  \frac{1}{\eta_{\lambda}}\frac{1}{\epsilon^2+1}+\frac{1}{\eta_{z}}\frac{1}{\xi^2+1}\right).
\end{equation}
Without a magnetic field, the last term on the RHS tends to be a Dirac$\delta$-function, representing the BIC obtained for $\eta_{z}=V_{z}=0$ case.

\begin{figure}[t]
\centering
\includegraphics[width=.75\linewidth]{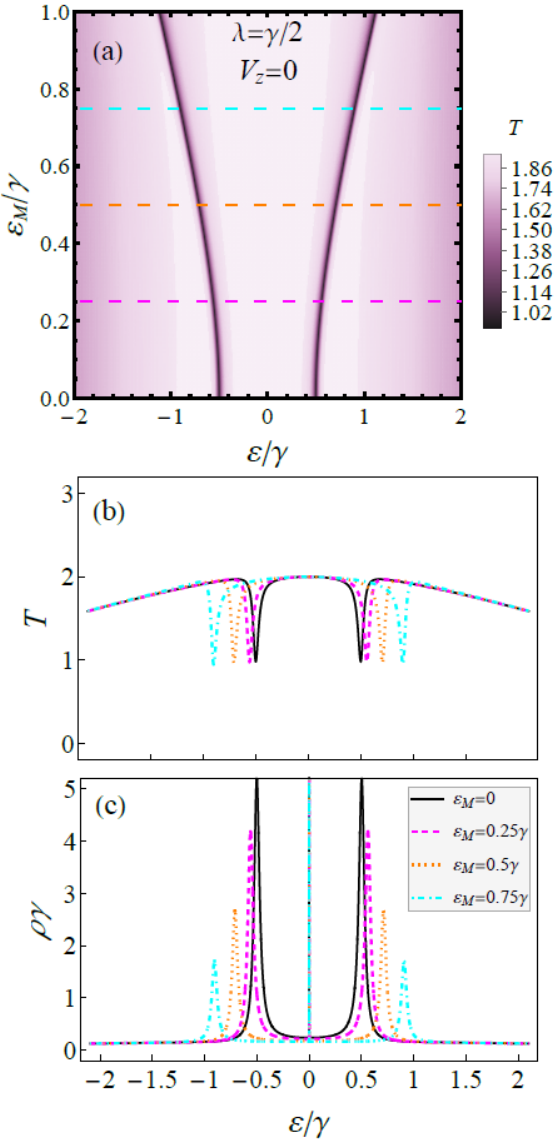}
\caption{(a) Color map of transmission probability $T$ as function of both $\ve$ and $\ve_{M}$. (b) Transmission probability as a function of $\ve$ for the fixed $\ve_{M}$ values indicated with dashed lines in panel (a). (c) The dimensionless density of states $\rho$ as a function of $\ve$ for the same fixed $\ve_{M}$ as above. In all panels we used $r=1$, $\lambda=\gamma/2$, and $V_{z}=\ve_{d}=0$.}\label{fig4}
\end{figure}

\begin{figure}[h!]
\centering
\includegraphics[width=.75\linewidth]{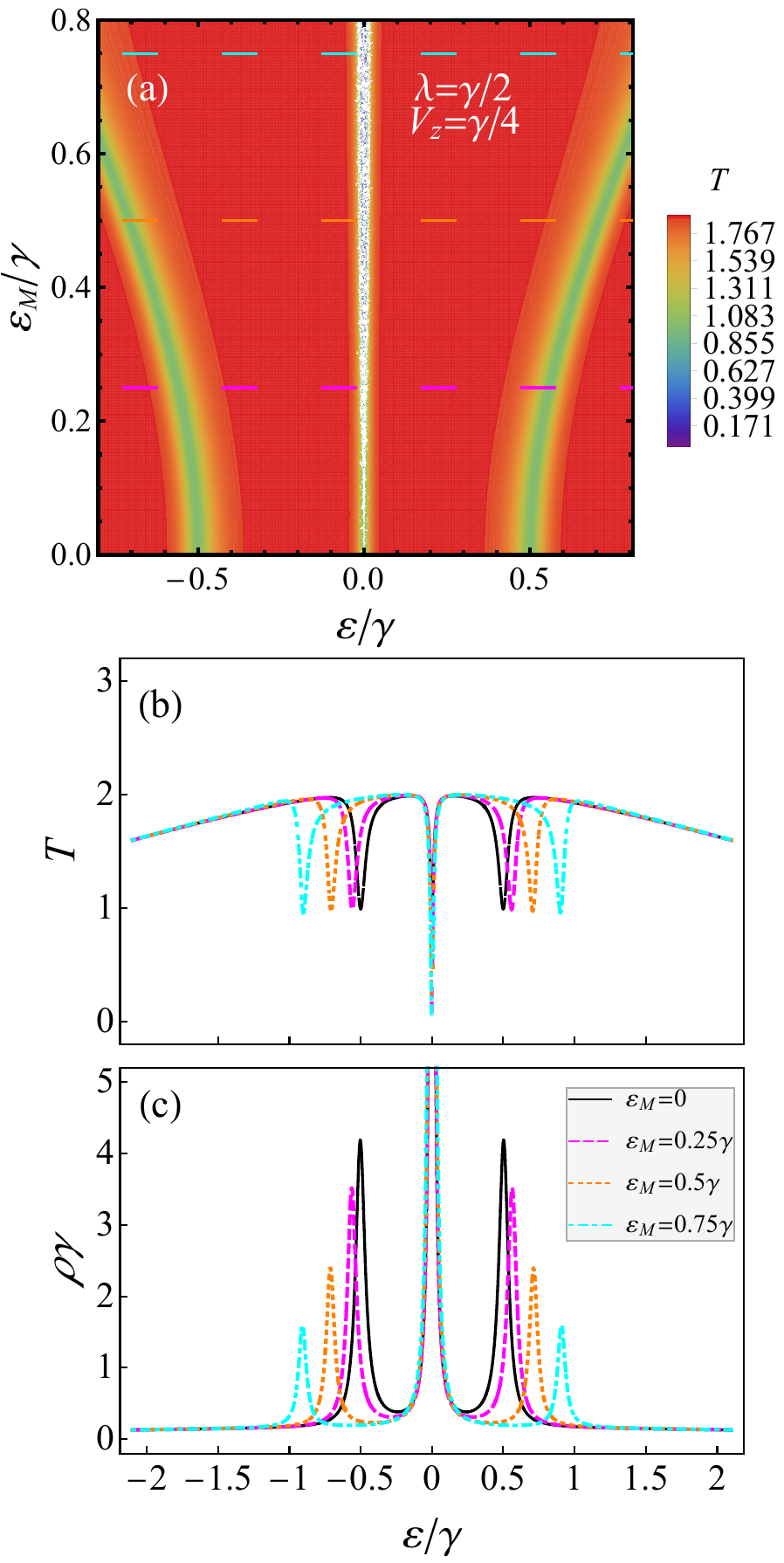}
\caption{(a) Color map of transmission probability $T$ as function of both $\ve$ and $\ve_{M}$. (b) Transmission probability as a function of $\ve$ for the fixed $\ve_{M}$ values indicated with dashed lines in panel (a). (c) The dimensionless density of states $\rho$ as a function of $\ve$ for the same fixed $\ve_{M}$ as above. In all panels we used $V_{z}=\gamma/4$, $r=1$, $\lambda=\gamma/2$, and $\ve_{d}=0$.}\label{fig5}
\end{figure}

Figure\ \ref{fig4}(a) shows the contour plot of the transmission probability $T$ as a function of $\varepsilon$ and $\varepsilon_M$, for fixed $V_{z}=\ve_{d}=0$ and $\lambda=\gamma/2$. No projection in the transmission is observed regardless of the value of $\ve_{M}$, as is presented complementary in Fig.\ref{fig4}(b). Nevertheless, the side non-vanishing antiresonances are placed at energies $\ve_{\pm}$ that evolves with $\ve_{M}$ as $\ve_{\pm}=\sqrt{\lambda^{2}+\ve_{M}^{2}}$. Besides, in Fig.  \ref{fig4}(c), we can observe that the BIC remained pinned at $\ve=0$, exhibiting a Dirac $\delta$-function in the DOS.

In Fig.~\ref{fig5}(a), we display the color map of the transmission probability $T$ as a function of $\ve$ and $\ve_{M}$ for fixed $\ve_{d}=0$, $\lambda=\gamma/2$ and $V_{z}=\gamma/4$. It is important to note that, regardless of the value of $\varepsilon_{M}$ considered, an antiresonance centered at $\varepsilon=0$ is observed, reaching zero whenever $\varepsilon_{M}\neq 0$ and attaining a finite value for $\varepsilon_{M}=0$, as additionally depicted in Fig.~\ref{fig5}(b). The non-vanishing antiresonance observed for $\varepsilon_{M}=0$ reach $T(\ve=\ve_{M}=0,V_{z}\neq0)=1$ ($T(\ve=\ve_{M}=0,V_{z}=0)/2$), exhibiting the leakage behavior of the MBS into the QD \cite{vernek2014subtle}. From Fig.~\ref{fig5}(c), illustrating the corresponding DOS, it is evident that the non-vanishing magnetic field, $V_{z}\neq0$, removes the BIC that existed at $\varepsilon=0$, regardless of the coupling parameter values. Lastly, it is noteworthy that the behavior of the antiresonances corresponding to the side states is similar to that described for the case without a magnetic field ($V_{z}=0$).

\begin{figure}[t]
\centering
\includegraphics[width=0.475\textwidth]{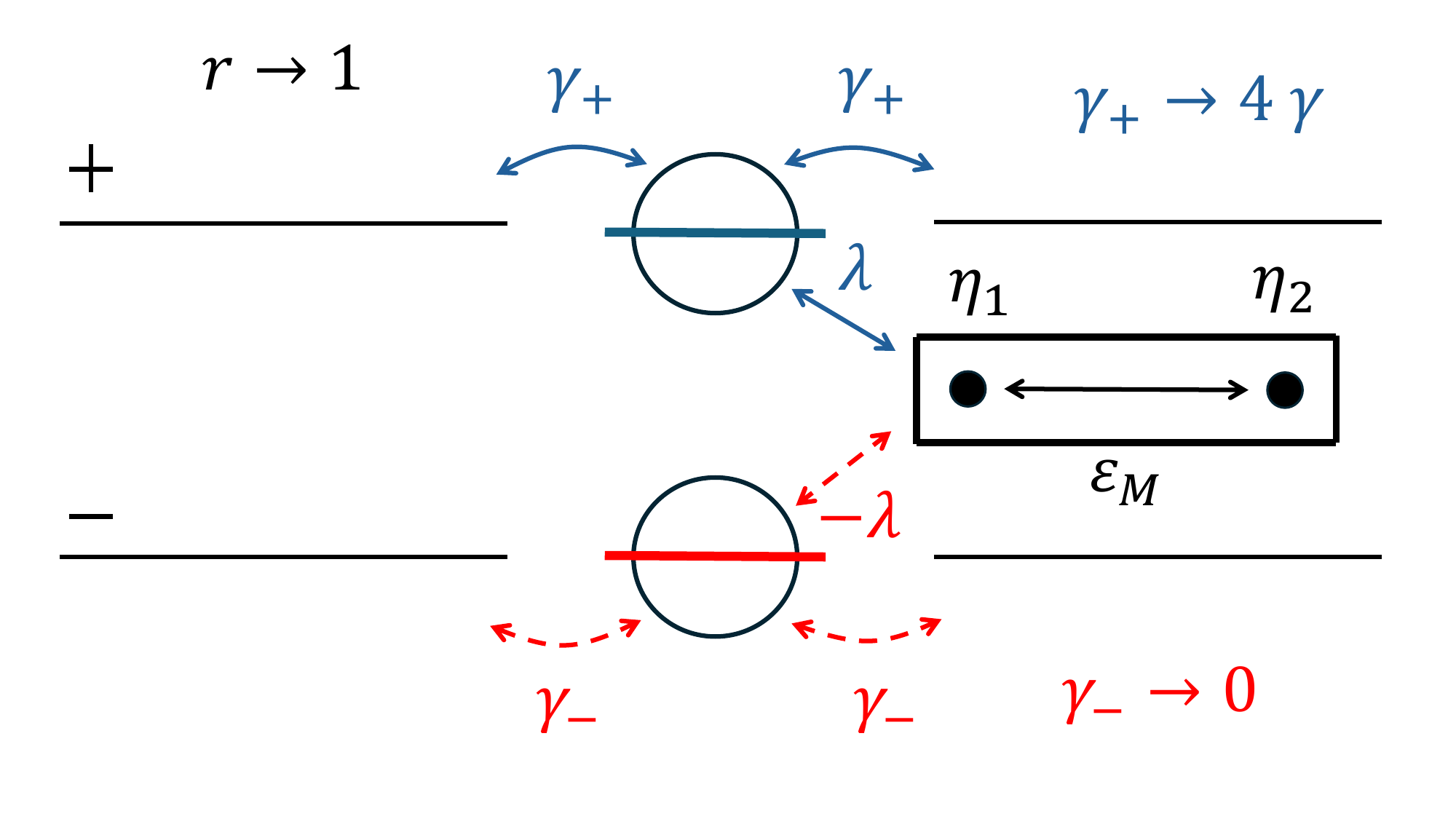}
\caption{Scheme of the symmetric and antisymmetric channels for the case $r=1$, $V_z =\ve_{d} = 0$}\label{fig6}
\end{figure}

From the aforementioned plots, it is clear that the BIC at zero energy remains robust under the coupling of the MBS to the spin-down projection in the QD. On the other hand, this state is coupled to the continuum when a magnetic field is turned on. This can be understood by analyzing the effective symmetric and antisymmetric channels that are present in this system and are obtained based on the transformation given in Eqs. (\ref{eq:greenpp}) to (\ref{eq:greenmp}). The effective system is depicted in FIG.\ \ref{fig6} in the case without a magnetic field. The symmetric state in the QD couples to the $+$channel through an amplitude $\gamma_{+}$ while the antisymmetric state is coupled to the $-$channel through an amplitude $\gamma_{-}$ that tends to $0$ when $r\rightarrow 1$. That means the antisymmetric state uncouples to the continuum, becoming a BIC. The inclusion of the MBS gives a channel that couples the $-$ states to the $+$ state, but the path that the electron should take from the $-$ state to the $+$ state is to pass to the $\eta_{1}$ state and then to the $+$ state. From the diagram, it is seen that this process includes a coupling of $-\lambda$ followed by a $+\lambda$, which would accumulate a dephase of $\pi$ in the wave function,  producing a destructive interference and canceling the leak of the antisymmetric state to the continuum.

\section{Summary}\label{summary}
In this article, we investigate the Fano-Rashba effect in a QD connected to a TSC nanowire that hosts two MBSs at its ends. We utilize the two-channel Fano-Anderson model to describe the system, including Rashba interaction and coupling to the TSC.
To resolve the problem, we use Green's function formalism and the equation of motion procedure. Our results of the transmission probability spectrum and the corresponding DOS in the QD demonstrate the emergence of Fano lineshapes due to the Rashba interaction and the coupling with the TSC. 

In addition, our results show that the BIC obtained for the symmetric coupling case ($r=1$) remains robust when $\ve_{M}=0$ in the presence of the MBSs for $V_{z}=0$. Besides, they demonstrate that the MBSs are inclined to leak into the QD when considering $V_{z}\neq0$.
The interference pattern is modified due to the interaction between the BIC and the MBSs in the system, impacting electronic transmission through the QD.
The latter could be accessible through low-temperature electronic conductance measurements.

\acknowledgments
J.P.R.-A is grateful for the financial support of FONDECYT Iniciaci\'on grant No. 11240637. P.A.O. acknowledges support from FONDECYT Grants No. 220700 and  1230933. B.G. acknowledges the support of ANID Becas/Doctorado Nacional 21242572.

\appendix

\section{\\Elements of the retarded GF in energy domain}\label{AppA}

In this appendix, we give a brief description of the procedure followed to obtain each element of the GF of the system.  We also presented the functions obtained.

The GF of the system is obtained using the equation of motion formalism. By definition, the GF in time domain, \(G_{\sigma, \sigma'}^{r}(t)\), is expressed by,

\begin{equation}
    G_{\sigma, \sigma'}^{r}(t) = -\frac{i}{\hbar} \Theta(t) \la\{ d_{\sigma}(t), d^{\dag}_{\sigma'}(0) \}\ra.
\end{equation}

Deriving with respect to time and multiplying by \(i \hbar\), we obtain the equation,

\begin{equation}
    i \hbar \frac{\partial G_{\sigma, \sigma'}^{r}(t)}{\partial t} = \delta(t) \delta_{\sigma , \sigma'} + \Theta(t) \left\langle\left\{ \frac{\drm}{\drm t} d_{\sigma}(t) , d_{\sigma'}^{\dag}(0) \right\}\right\rangle.
    \label{eq:eom_qdot}
\end{equation}

The time derivative of \(d_{\sigma}(t)\) can be calculated using its equation of motion,

\begin{equation}
    \frac{\drm}{\drm t} d_{\sigma}(t) = \frac{1}{i\hbar} \left[ d_{\sigma} , H \right].
\end{equation}

Then, the Eq.\ \eqref{eq:eom_qdot} can be written as

\begin{widetext}

\begin{equation}
    i \hbar \frac{\partial G_{\sigma, \sigma'}^{r}(t)}{\partial t} = \delta(t) \delta_{\sigma,\sigma'} + \ve_{\sigma} G_{\sigma, \sigma'}^{r}(t) - \frac{\lambda}{\sqrt{2}} \delta_{\sigma, \down} G_{f,\sigma'}^{r}(t) - \frac{\lambda}{\sqrt{2}} \delta_{\sigma, \down} F_{f,\sigma'}^{r}(t) + \sum_{\mathbf{k},\alpha} V_{k,\alpha, \sigma} G_{\mathbf{k},\alpha, \sigma , \sigma'}^{r}(t) + \sum_{\mathbf{k},\alpha} W G_{\mathbf{k},\alpha, \bar{\sigma} , \sigma'}^{r}(t).
    \label{eq:eom_qdot_time}
\end{equation}

\end{widetext}

Where the next GF's were defined,
\begin{align}
    G_{f, \sigma'}^{r}(t) &= -\frac{i}{\hbar} \Theta(t) \la\{ f_{\sigma}(t), d^{\dag}_{\sigma'}(0) \}\ra.\\
    G_{\mathbf{k},\alpha, \sigma , \sigma'}^{r}(t) &= -\frac{i}{\hbar} \Theta(t) \la\{ c_{\mathbf{k},\alpha, \sigma}(t), d^{\dag}_{\sigma'}(0) \}\ra.
\end{align}
It is necessary to perform the same procedure to all the GF that appear in Eq.\ \eqref{eq:eom_qdot_time}. After doing so, we obtain a system of differential equations for the GF in the time domain. In the energy domain, the equations for the GF become a set of linear equations. In the latter scenario, the solution of this set gives the elements of the GF presented in Eq.\ (\ref{greenmatriz}), which are given by the following sixteen expressions.
\begin{align}
    G_{\up,\up}^{r}(\ve) &= \frac{1}{ \ve - \ve_{d \up} - \Sigma_{1}(\ve,\up) - K_{3}(\ve,\ve_{\up},\ve_{\down}) \Sigma_{2}(\ve,\up)}\,, \\
    G_{\down,\down}^{r}(\ve) &= \frac{1}{K_{2}(\ve,\ve_{\up},\ve_{\down})}\,, \\
    G_{\up,\down}^{r}(\ve) &= \frac{\Sigma_{2}(\ve,\up)}{K_{2}(\ve,\ve_{\up},\ve_{\down})[\ve - \ve_{d,\up} - \Sigma_{1}(\ve,\up)]}\,,\\
    G_{\down,\up}^{r}(\ve) &= \frac{K_{3}(\ve,\ve_{\up},\ve_{\down})}{ \ve - \ve_{d\up} - \Sigma_{1}(\ve,\up) - K_{3}(\ve,\ve_{\up},\ve_{\down}) \Sigma_{2}(\ve,\up) }\,,
\end{align}
\begin{align}
    F_{\up,\up}^{r}(\ve) &= \frac{-\lambda^{2}K_{0}(\ve)\tilde{\Sigma}_{2}(\ve,\up) G_{\down,\up}^{r}(\ve)}{N(\ve,\ve_{\up},\ve_{\down})}\,,\\
     F_{\down,\down}^{r}(\ve) &= \frac{-\lambda^{2} K_{0}(\ve)[\ve+\ve_{\up}-\tilde{\Sigma}_{1}(\ve,\up)]G_{\down,\down}^{r}(\ve)}{N(\ve,\ve_{\up},\ve_{\down})}\,,\\
    F_{\up,\down}^{r}(\ve) &= \frac{-\lambda^{2}K_{0}(\ve) \tilde{\Sigma}_{2}(\ve,\up) G_{\down,\down}^{r}(\ve)}{N(\ve,\ve_{\up},\ve_{\down})}\,,\\
    F_{\down,\up}^{r}(\ve) &= \frac{-\lambda^{2} K_{0}(\ve)[\ve+\ve_{\up}-\tilde{\Sigma}_{1}(\ve,\up)]G_{\down,\up}^{r}(\ve)}{N(\ve,\ve_{\up},\ve_{\down})}\,,
\end{align}
\begin{align}
    \mathcal{G}_{\up,\up}^{r}(\ve) &= \frac{1}{ \ve + \ve_{d \up} - \tilde{\Sigma}_{1}(\ve,\up) - K_{3}(\ve,-\ve_{\up},-\ve_{\down}) \tilde{\Sigma}_{2}(\ve,\up)}\,, \\
    \mathcal{G}_{\down,\down}^{r}(\ve) &= \frac{1}{K_{2}(\ve,-\ve_{\up},-\ve_{\down})}\,, \\
    \mathcal{G}_{\up,\down}^{r}(\ve) &= \frac{\tilde{\Sigma}_{2}(\ve,\up)}{K_2(\ve,-\ve_{\up},-\ve_{\down})[\ve + \ve_{d,\up} - \tilde{\Sigma}_{1}(\ve,\up)]}\,,\\
    \mathcal{G}_{\down,\up}^{r}(\ve) &= \frac{K_{3}(\ve,-\ve_{\up},-\ve_{\down})}{ \ve + \ve_{d\up} - \tilde{\Sigma}_{1}(\ve,\up) - K_{3}(\ve,-\ve_{\up},-\ve_{\down}) \tilde{\Sigma}_{2}(\ve,\up) }\,,
\end{align}
\begin{align}
    \mathcal{F}_{\up,\up}^{r}(\ve) &= \frac{-\lambda^{2}K_{0}(\ve)\Sigma_{2}(\ve,\up) \mathcal{G}_{\down,\up}^{r}(\ve)}{N(\ve,-\ve_{\up},-\ve_{\down})}\,,\\
    \mathcal{F}_{\down,\down}^{r}(\ve) &= \frac{-\lambda^{2} K_{0}(\ve)[\ve-\ve_{\up}-\Sigma_{1}(\ve,\up)]\mathcal{G}_{\down,\down}^{r}(\ve)}{N(\ve,-\ve_{\up},-\ve_{\down})}\,,\\
    \mathcal{F}_{\up,\down}^{r}(\ve) &= \frac{-\lambda^{2}K_{0}(\ve) \Sigma_{2}(\ve,\up) \mathcal{G}_{\down,\down}^{r}(\ve)}{N(\ve,-\ve_{\up},-\ve_{\down})}\,,\\
    \mathcal{F}_{\down,\up}^{r}(\ve) &= \frac{-\lambda^{2} K_{0}(\ve)[\ve-\ve_{\up}-\Sigma_{1}(\ve,\up)]\mathcal{G}_{\down,\up}^{r}(\ve)}{N(\ve,-\ve_{\up},-\ve_{\down})}\,,
\end{align}
where we have defined the following functions
\begin{widetext}
\begin{align}
    K_{0}(\ve) &= \frac{\ve}{\ve^{2} - \ve_{M}^{2}}\,, \\
    K_{1}(\ve,\ve_{\up},\ve_{\down}) &= \frac{-\lambda^{2}K_{0}(\ve)}{\ve + \ve_{d \down} - \tilde{\Sigma}_{1}(\ve,\down) - \lambda^{2} K_{0}(\ve) + \dfrac{\tilde{\Sigma}_{2}(\ve,\down)\tilde{\Sigma}_{2}(\ve,\up)}{\ve+\ve_{d\up} - \tilde{\Sigma}_{1}(\ve,\up)}}\,, \\
   K_{2}(\ve,\ve_{\up},\ve_{\down}) &= \ve - \ve_{d\down} - \Sigma_{1}(\ve,\down) - \lambda^{2} K_{0}(\ve) + \lambda^{2} K_{0}(\ve)K_{1}(\ve,\ve_{\up},\ve_{\down}) - \frac{\Sigma_{2}(\ve,\down)\Sigma_{2}(\ve,\up)}{\ve - \ve_{d,\up} - \Sigma_{1}(\ve,\up)}\,, \\
   K_{3}(\ve,\ve_{\up},\ve_{\down}) &= \frac{\Sigma_{2}(\ve,\down)}{\ve - \ve_{d\down} - \Sigma_{1}(\ve,\down) - \lambda^{2}K_{0}(\ve) K_{1}(\ve)}\,,\\
    N(\ve,\ve_{\up},\ve_{\down}) &= [\ve + \ve_{d\down} - \tilde{\Sigma}_{1}(\ve,\down) - \lambda^{2} K_{0}(\ve)][\ve + \ve_{d \up} - \tilde{\Sigma}_{1}(\ve,\up)]-\tilde{\Sigma}_{2}(\ve,\up) \tilde{\Sigma}_{2}(\ve,\down)\,,
\end{align}
\end{widetext}
where the self-energies are defined by
\begin{align*}
    \Sigma_{1}(\ve,\sigma) &= \sum_{\mathbf{k},\alpha}\left( \frac{\lvert V_{\mathbf{k},\alpha,\sigma}\lvert^{2}}{\ve - \ve_{\mathbf{k},\sigma}} + \frac{\lvert t_{R} \lvert^{2}}{\ve - \ve_{\mathbf{k}, \bar{\sigma}}} \right)\,, \\
    \Sigma_{2}(\ve,\sigma) &= \sum_{\mathbf{k},\alpha}\left( \frac{V_{\mathbf{k},\alpha,\sigma} t^{*}_{R}}{\ve - \ve_{\mathbf{k},\sigma}} + \frac{t_{R} V^{*}_{\mathbf{k},\alpha,\bar{\sigma}}}{\ve - \ve_{\mathbf{k},\bar{\sigma}}} \right)\,,\\
    \tilde{\Sigma}_{1}(\ve,\sigma) &= \sum_{\mathbf{k},\alpha}\left( \frac{\lvert V_{\mathbf{k},\alpha,\sigma}\lvert^{2}}{\ve + \ve_{\mathbf{k},\sigma}} + \frac{\lvert t_{R} \lvert^{2}}{\ve + \ve_{\mathbf{k}, \bar{\sigma}}} \right)\,,\\
    \tilde{\Sigma}_{2}(\ve,\sigma) &= \sum_{\mathbf{k},\alpha}\left( \frac{V^{*}_{\mathbf{k},\alpha,\sigma} t_{R}}{\ve + \ve_{\mathbf{k},\sigma}} + \frac{t^{*}_{R} V_{\mathbf{k},\alpha,\bar{\sigma}}}{\ve + \ve_{\mathbf{k},\bar{\sigma}}} \right)\,.
\end{align*}

\twocolumngrid

\bibliographystyle{apsrev4-1}
\bibliography{biblio}

\end{document}